\documentclass[twocolumn,secnumarabic,amssymb, nobibnotes, superscriptaddress,aps, prd]{revtex4-1}

\newcommand{\bea}{\begin{eqnarray}}
\newcommand{\ena}{\end{eqnarray}}
\newcommand{\be}{\begin{equation}}
\newcommand{\ee}{\end{equation}}
\newcommand{\beann}{\begin{eqnarray*}}
\newcommand{\enann}{\end{eqnarray*}}

\usepackage{eurosym}
\usepackage{graphicx}
\usepackage{amsmath}
\usepackage{mathrsfs}
\usepackage{bm}
\usepackage{color}
\usepackage{amsmath}
\usepackage{amssymb}

\usepackage{multirow}
\usepackage{array}
\usepackage{subfig}

\usepackage[utf8]{inputenc}
\usepackage{graphics,graphicx}
\usepackage{amsmath,amssymb,amsfonts,latexsym,cancel}
\usepackage{amsmath,amssymb}

\begin{document}

\title{Tidal Deformation and Radial Pulsations of Neutron Star with Holographic Multiquark Core}

\author{Sitthichai Pinkanjanarod}
\email{quazact@gmail.com, Sitthichai.P@student.chula.ac.th}
\affiliation{High Energy Physics Theory Group, Department of Physics,Faculty of Science, Chulalongkorn University, Bangkok 10330, Thailand}
\affiliation{  Department of Physics, Faculty of Science, Kasetsart University, Bangkok 10900, Thailand}

\author{Piyabut Burikham}
\email{piyabut@gmail.com}
\affiliation{High Energy Physics Theory Group, Department of Physics,Faculty of Science, Chulalongkorn University, Bangkok 10330, Thailand}

\author{Supakchai Ponglertsakul}
\email{supakchai.p@gmail.com}
\affiliation{High Energy Physics Theory Group, Department of Physics,Faculty of Science, Chulalongkorn University, Bangkok 10330, Thailand}
\affiliation{ Department of Physics, Faculty of Science, Silpakorn University,Nakhon Pathom 73000,
	Thailand}

\date{\today}

\begin{abstract}

Tidal deformation of neutron star with multiquark core is calculated using nuclear and multiquark equations of state.  The equation of state of the multiquark phase from the holographic Sakai-Sugimoto~(SS) model is relatively stiff in the low density region and becomes softer at high densities. The values of Love number and dimensionless deformation parameter, $k_{2}$ and $\Lambda$, are found to be within the physically viable range under the present constraints.  Radial pulsation frequencies of the multiquark core for $n=0-5$ modes are calculated for the entire mass range.  For $M_{\rm core}\simeq 2 M_{\odot}$, the fundamental-mode frequency is approximately $2.5$ kHz for the energy density scale $\epsilon_{s}=23.2037$ GeV/fm$^{3}$ of the holographic SS model, this frequency is proportional to $\sqrt{\epsilon_{s}}$.
	
\end{abstract}
\maketitle

\section{Introduction}\label{sec-introd}

A number of detections of massive neutron stars~(NS) with masses approximately larger than 2 solar mass~($M_{\odot}$)~\cite{Antoniadis:2013pzd,Miller:2021qha,Clark:2002db,Romani:2012rh,Romani:2012jf,vanKerkwijk:2010mt,Linares:2018ppq,Bhalerao:2012xe,Nice:2005fi,Demorest:2010bx,Freire:2007sg,Quaintrell:2003pn} indicates that the equation of state~(EoS) of the core of the massive NS could be softer than the usual stiff baryonic nuclear matter in the crust. Substantial scan of various forms of the EoS based on the sound speed and adiabatic index reveals the possibility that the massive NS has a quark matter core~\cite{Annala:2019puf}.    

The equation of state of nuclear matter at high density is still an open problem of the strong interaction. For low density and extremely high temperature~(around trillion K), interaction between quarks and gluons described by quantum chromodynamics (QCD) is weak and the quarks and gluons could become deconfined in a phase of quark-gluon plasma~(QGP).  However, as the NS cooled down, QCD becomes strongly coupled at low temperature while the baryon chemical potential/density could still be large.  In particular, at the core of the NS, the radius of confinement could become larger due to extreme density and quarks could be effectively deconfined from the localized hadrons.  However, the quarks are yet to be confined by gravity within the star.  There is a number of nonperturbative QCD approaches for the EoS of the interior of the NS. For instance, the bag models of free quarks inside the star are simple and effective but they are based on the assumption of weakly coupled quarks. Lattice QCD, which is very useful for nonperturbative QCD, however becomes unreliable at large baryon chemical potential~\cite{Lombardo:2006yc}. Colour superconductivity~(CSC) assumes that quarks form diquark condensate at high density which can be described by the Nambu-Jona-Lasinio type theory~\cite{Alford:2007xm}. 

There is a complementary powerful tool for studying strongly coupled gauge theories that emerged from the discovery of the gauge/gravity duality, i.e., the  AdS/CFT correspondence~\cite{Gubser:1998bc,maldacena,witb}. One of the most astonishing features in the gauge/string duality is the two theories defined in different dimensions are conjectured to be equivalent under a strong-weak duality that is useful to study the strongly coupled QCD. In the holographic QCD the mesons and baryons are dually described by higher dimensional configurations consisting of open strings and D-branes in the specific background~(see e.g. Ref.~\cite{Witten:1998xy,Callan:1998iq,Brandhuber:1998xy} for earlier works and Ref.~\cite{Nakas:2020hyo} for more recent development).  In Ref.~\cite{bch}, holographic multiquark~(MQ) phase in the deconfined background with broken chiral symmetry based on the Sakai-Sugimoto model~\cite{ss1, ss2,Aharony_chiral,Bergman:2007wp} is demonstrated to be thermodynamically prefered than the QGP phase at high chemical potential/density for low to moderate temperatures up to the QGP phase transition around trillion Kelvins. In Ref.~\cite{bhp}, the EoS of the multiquark phase is derived and applied to the hypothetical multiquark star and NS with multiquark core. Preliminary estimation allows the existence of NS with multiquark core with masses as large as $3 M_{\odot}$.

With the prospect of massive NS above $2M_{\odot}$ containing quark matter core from observations, it is interesting to investigate how the holographic multiquark EoS fit into the observational constraints. In addition to the mass $M$ and radius $R$, the ratio $M/R$ which represents compactness of the star also plays a crucial role in determination of the exact EoS of the matter inside the NS and possibly other kinds of extreme compact objects such as quark stars. A good probe for the compactness is deformability of the star parametrized by Love number $k_{2}$~\cite{Mora:2003wt,Berti:2007cd,Flanagan:2007ix,Hinderer:2007mb} and the dimensionless parameter $\Lambda$ defined in Sect.~\ref{SectLove}.  Tidal deformabilities and other properties such as moment of inertia, sound speed and EoS of hybrid stars in holographic models of quark and nuclear matter are also calculated in Ref.~\cite{Annala:2017tqz,Jokela:2018ers,Zhang:2019tqd,Jokela:2020piw}.  In this work we calculate the Love number $k_{2}$ and $\Lambda$ of the NS with multiquark core by adapting Chiral Effective Field Theory~(CET) EoS~\cite{Tews:2012fj} and FYSS EoS~\cite{FYSS_2017,FYSS_2013,FYSS_2011} for the nuclear crust.  The deformation parameters of the multiquark star~(MQS) are also calculated for comparison. We found that for massive NS with multiquark core, the values of deformation parameters are within the allowed constraints from LIGO/Virgo. Radial pulsation frequencies of the fundamental and excited modes of the MQ core are calculated and the zeroth-mode instability is verified to occur at the maximum masses. 

This work is organized as the following. Section~\ref{sec-holomq} reviews the holographic multiquark model and the corresponding EoS.  The adiabatic index and sound speed of multiquark core are explored in Section~\ref{secgam}. The tidal deformation of the NS with multiquark core and MQS are calculated in Section~\ref{SectLove}.  The radial pulsations of the stars are discussed in Section~\ref{secpul}.  Section~\ref{sec-con} concludes the work. 

\section{Holographic multiquark and the EoS}\label{sec-holomq}
According to gauge-gravity duality from superstring theories, bound states of multiquarks can be described holographically by strings and branes in D4/D8/$\overline{\text{D8}}$ system called Sakai-Sugimoto (SS) model~\cite{ss1,ss2}. The SS model can capture most features of the QCD e.g. low-energy spectrum of mesons, confinement/deconfinement transition and chiral symmetry breaking. In this model, hadrons naturally exist in the confined phase however, another kind of bound states of quarks can also exist in the deconfined phase at low and intermediate temperatures when the chemical potential/density is sufficient large i.e., the multiquark states~\cite{bch,bhp}~(see e.g. Ref.~\cite{Burikham:2011zz} for a concise review of holographic multiquarks).

\subsection{Equation of state}  \label{sec-eos}
As explored in details in Ref.~\cite{bhp}, the equations of state for the multiquark involving pressure $P$ and mass density $\rho$ are given in terms of the number density $n$ which could be summarized in the dimensionless form as
\bea
P &=& a n^{2}+b n^{4},  \notag \\
\rho  &=& \mu_{0}n+a n^{2}+\frac{b}{3}n^{4},  \label{eosmq1}
\ena
for small $n$~(``mql'') and 
\bea
P &=& k n^{7/5},  \notag \\
\rho  &=& \rho_{c} +\frac{5}{2}P+\mu_{c}\left(n-n_{c}\right) \notag \\
&&+kn_{c}^{7/5}-\frac{7k}{2}n_{c}^{2/5}n,  \label{eosmq2}
\ena
for large $n$~(``mqh'') respectively.  Note that all quantities are dimensionless and the conversion table to physical quantities is given in the Appendix. The parameter $\mu_{0}$ is the onset chemical potential of the multiquark phase.  The critical mass density, number density and chemical potential where the EoS changes from large to small $n$ are denoted by $\rho_{c}, n_{c}$ and $\mu_{c}$ respectively.  EoS of the deconfined multiquark matter also depends on the colour charge of the multiquark which is parametrized by $n_{s}$, see Ref.~\cite{bch,bhp} for more details.  For $n_{s}=0, \rho_{c}=0.0841077, n_{c}=0.215443, \mu_{c}=0.564374~(\text{mqh}), a=1, b=0, \mu_{0}=0.17495~(\text{mql})$. For $n_{s}=0.3, \rho_{c}=0.0345996, n_{c}=0.086666, \mu_{c}=0.490069~(\text{mqh}), a=0.375, b=180.0, \mu_{0}=0.32767~(\text{mql})$.  In both cases of $n_{s}=0, 0.3$, we found unique value $k=10^{-0.4}$ which reflects universal behaviour of the MQ at high density regardless of the colour charges.  The SS model has only two free parameters, the number fraction of hanging strings representing the colour charges $n_{s}$ and the energy density scale $\epsilon_{s}$ defined in Ref.~\cite{bch,bhp}.  The pressure and mass density scale with $\epsilon_{s}$ as $P, \rho \sim \epsilon_{s}$.  Consequently, the mass and radius of the pure MQS have the same scaling $M, R \sim \epsilon_{s}^{-1/2}$. Remarkably, the compactness $M/R$ is thus independent of $\epsilon_{s}$.

As the density decreases with increasing radial distance from the center of the star, the confinement phase transition occurs.  The gluons and multiquarks undergo a transition into the colour-singlet baryonic matter.  As in Ref.~\cite{Pinkanjanarod:2020mgi}, we use the CET stiff EoS for the baryonic crust of the NS.  Moreover, the FYSS EoS~\cite{FYSS_2017,FYSS_2013,FYSS_2011} at temperature $T=0.1$ MeV is also implemented as an alternative crust EoS for comparison.  The exact values of $(P, \mu)$ at the confinement phase transition can be found by studying the $P-\mu$ diagram of the two phases, the star profile as well as the full mass-radius relations are obtained, see Ref.~\cite{Pinkanjanarod:2020mgi} for details.  In this work, we will focus on the deformability and dynamical pulsations of such hybrid stars and the MQS.

\section{Adiabatic indices and sound speed}  \label{secgam}

The adiabatic indices describe how pressure of the system changes with the density.  The content of a star could either have the stiff EoS where the adiabatic index is large or soft EoS when it is small.  Large adiabatic index of matter in the interior could make the star more massive due to the larger sustaining pressure per mass.  Additionally, it represents compressibility of the matter inside the star which determines the pulsation frequencies and dynamical stability of the star under perturbation~(see Eq.~(\ref{Delxi})).   

Specifically, one kind of adiabatic index could be defined with respect to the number density $n$ as
\begin{eqnarray}
\Gamma &=& \frac{n}{P}\left(\frac{dP}{dn}\right).  \label{Gamdef}
\end{eqnarray}
This form of adiabatic index (\ref{Gamdef}) reflects the fact that in a pulsating star, the number density $n$ varies with the pulsation. However, when considering the stiffness of the EoS, we shall refer to the conventional relativistic adiabatic~(polytropic) index of the form, $\gamma = \displaystyle{ \frac{\rho}{P}(\frac{dP}{d\rho})}$.  For massive NS with multiquark core at the {\it maximum mass}, the range of values of $(\Gamma, \gamma, c_{s}^{2})$ within the star are given in Table~\ref{tabgam}.  The high-density multiquark EoS is apparently softer than the low-density EoS with smaller $\gamma$.  The low-density ``mql'' EoS is considered stiff since $\gamma > 5/3$ while the ``mqh'' EoS in the core is softer with $\gamma \simeq 1.1$. 
\begin{table}
\footnotesize
\begin{tabular}{ | c | c | c | }
\hline
&\multicolumn{2}{c|}{}\\
& \multicolumn{2}{c|}{$(\Gamma, \gamma, c_s^2)$} \\
$\epsilon_{s}$  & \multicolumn{2}{c|}{} \\
\cline{2-3}
(GeVfm$^{-3}$)&& \\
& mqh & mql~(small to large $r$) \\
&& \\
\hline
23.2037&&\\
\& FYSS, & (1.4, 1.1, 0.43) & $(3.6 - 2.3, 2.6 - 2.2, 0.97 - 0.05)$\\
$Y = 0.10$ &&\\
\hline
23.2037&&\\
\& FYSS, & (1.4, 1.1, 0.43) & $(3.6 - 2.3, 2.6 - 2.2, 0.97 - 0.06)$\\
$Y = 0.14$ &&\\
\hline
&&\\
26 \& CET & (1.4, 1.1, 0.43) & $(3.6 - 3.1, 2.6 - 2.8, 0.97 - 0.33)$\\
&&\\
\hline
&&\\
28 \& CET & (1.4, 1.1, 0.43) & $(3.6 - 3.0, 2.6 - 2.7, 0.97 - 0.26)$\\
&&\\
\hline
\end{tabular}
\caption{Values of $(\Gamma, \gamma, c_s^2)$ for high density multiquark ``mqh" in the inner core and low density ``mql" in the outer core with CET/FYSS nuclear crust at the {\it maximum mass} of the massive NS with MQ core.}
\label{tabgam}
\end{table}

\section{Tidal Love number and tidal deformation}   \label{SectLove}

\subsection{Definitions of tidal Love number and tidal deformation}
Consider a static, spherically symmetric star of mass $M$ placed
in an external static quadrupole tidal field $\epsilon_{ij}$. The star will deform in response and obtain a quadrupole moment $Q_{ij}$.
In the star's local asymptotic rest frame, the metric component $g_{tt}$ at large $r$ is given by~\cite{Thorne:1997kt}
\begin{eqnarray}
-\frac{(1+g_{tt})}{2} =&-&\frac{M}{r}-\frac{3Q_{ij}}{2r^3}\left(n^i n^j-\frac{1}{3}\delta^{ij}\right)+O\left(\frac{1}{r^3}\right)\nonumber\\&+& \frac{1}{2}\epsilon_{ij}x^ix^j + O(r^3)
\label{gtt_perturb_comp}
\end{eqnarray}
where $n^i \equiv x^i/r$. We are interested in the strong field case. For an isolated body in a static situation, these moments are uniquely defined: $\epsilon_{ij}$ and $Q_{ij}$ are the coordinate-independent quantities.  Correspondingly, to the linear order in $\epsilon_{ij}$, the induced quadrupole will be given in terms of the tidal field by 
\begin{eqnarray}
Q_{ij} &=& -\lambda \epsilon_{ij},
\label{Q_epsilon}
\end{eqnarray}
where $\lambda$ is a constant associated with the $\ell=2$ tidal Love number $k_2$ by~\cite{Flanagan:2007ix}
\begin{eqnarray}
k_2 &=&\frac{3}{2}\lambda R^{-5}.
\label{k2_lambda}
\end{eqnarray}

The multipole moments $Q_{ij}$ and $\epsilon_{ij}$ can be decomposed as a series of spherical harmonic basis functions
\begin{eqnarray}
\epsilon_{ij} &=& \sum_{\text{m}=-2}^{2}\epsilon_{\text{m}}\mathcal{Y}^{2\text{m}}_{ij}, \notag
\label{epslon_exp}\\
Q_{ij} &=& \sum_{\text{m}=-2}^{2}Q_{\text{m}}\mathcal{Y}^{2\text{m}}_{ij}.  \notag
\label{Q_epsilon_exp}
\end{eqnarray}
The tensor spherical harmonic $\mathcal{Y}^{2\text{m}}_{ij}$, which are symmetric and traceless, are defined by
\begin{eqnarray}
Y_{2m}(\theta,\phi) &=& \mathcal{Y}^{2\text{m}}_{ij}n^i n^j,  \notag
\label{Y_exp}
\end{eqnarray}
where $Y_{2m}(\theta,\phi)$ is the spherical harmonic with $\ell =2$ and $n \equiv (\sin \theta \cos \phi, \sin \theta \sin \phi, \cos \theta)$. Thus, from Eqs.~(\ref{Q_epsilon}), the tidal deformation $\lambda$ for a particular mode $m$~($\ell =2$) can be written as
\begin{eqnarray}
\lambda &=& - \frac{\epsilon_{\text{m}}}{Q_{\text{m}}},  \notag
\label{lamda}
\end{eqnarray}
and in the dimensionless form as~\cite{Raithel:2018ncd},
\begin{eqnarray}
\Lambda \equiv \frac{\lambda}{M^5} \equiv \frac{2}{3} k_2 \left(\frac{R}{M}\right)^5.
\label{lamda_dimless}
\end{eqnarray}
Note that in this work, we will use the geometric units $G=c=1$ in our calculation of $k_{2}, \Lambda$.

\subsection{Calculations of tidal Love number and tidal deformation}

In the moderately strong field, e.g. long before the merging of the binary, we assume the perturbation is small comparing to the background metric $g^{(0)}_{\alpha \beta}$ which takes the standard spherically symmetric form
\begin{eqnarray}
g_{\alpha \beta} &=& g^{(0)}_{\alpha \beta} +  h_{\alpha \beta}, \notag \\
g^{(0)}_{\alpha \beta}&=& \text{diag}[-e^{\nu(r)},e^{\lambda(r)},r^{2},r^{2}\sin^{2}\theta],
\label{perturb_metric}
\end{eqnarray}
where $h_{\alpha \beta}$ is a linearized metric perturbation. We only consider the $\ell = 2$, static, even-parity perturbations in the Regge-Wheeler gauge~\cite{Regge:1957td,Hinderer:2009ca}. The linearized metric perturbation $h_{\alpha \beta}$ could then be expressed in the diagonal form as~\cite{Thorne_Campolattaro,Hinderer:2009ca}

\begin{eqnarray}
h_{\alpha \beta} &=  -\text{diag}[
e^{\nu(r)}H(r), e^{\lambda(r)}H(r),r^2 K(r),\nonumber\\ &r^{2}K(r)\text{sin}^2\theta ] Y_{2m}(\theta, \phi).\label{Lperturb}
\end{eqnarray}

\begin{figure}[h]
	\centering
	\includegraphics[width=0.48\textwidth]{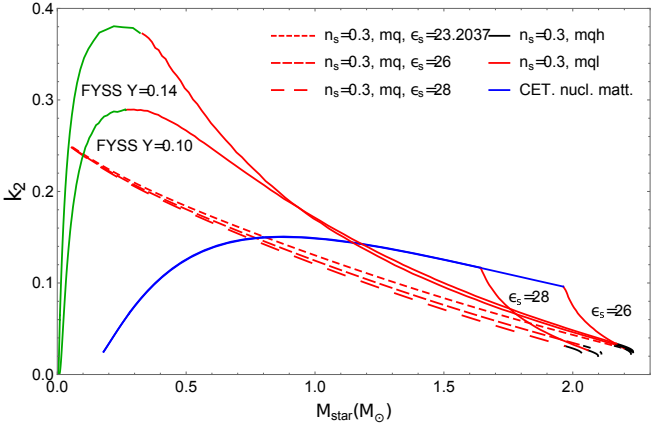}
	\caption{Tidal Love number $k_2$ versus mass of MQS/NS with MQ core.}
	\label{k2fig}
\end{figure}
\begin{figure}[h]
	\centering
	\includegraphics[width=0.48\textwidth]{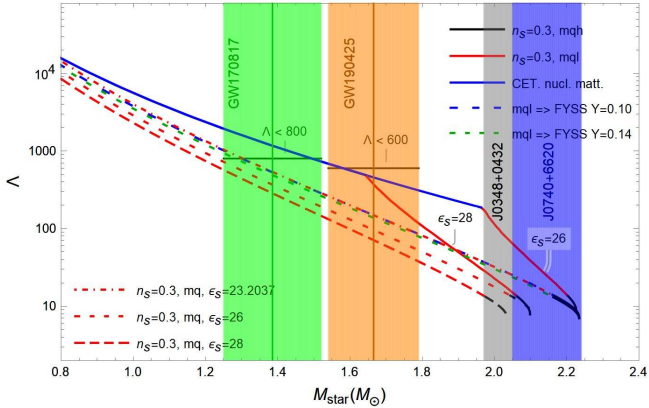}
	\caption{Dimensionless tidal deformation $\Lambda$ versus mass of MQS/NS with MQ core.}
	\label{lamMfig}
\end{figure}

The nonzero components of the perturbations of the stress-energy tensor are $\delta T ^0_0 = -\delta \rho = -(dP/d\rho)^{-1}\delta P$
and $\delta T ^i_i = \delta P$. Insert these relations and the metric perturbation Eq.~(\ref{Lperturb}) into the linearized Einstein equation  $\delta G_{\alpha}^{\beta} = 8\pi\delta T_{\alpha}^{\beta}$, and after elimination of $K$, we obtain the following differential equation~\cite{Hinderer:2009ca}
\begin{eqnarray}
H^{\prime \prime} 
&+& \frac{2}{r} \left(1-2\frac{m}{r}\right)^{-1}H^{\prime}\left\{1 -\frac{m}{r} - 2 \pi r^{2}(\rho-P)\right\}\nonumber\\
&+& 2\left(1-2\frac{m}{r}\right)^{-1} H\left\{2\pi \left(5 P + 9\rho + \frac{\rho+P}{dP/d\rho}\right)\right.\nonumber\\ 
&-& \left. \frac{3}{r^2} - 2\left(1-2\frac{m}{r}\right)^{-1} \left(\frac{m}{r^2}+4\pi r P\right)^2\right\} = 0, \label{H_perturb}
\end{eqnarray}
where $^\prime$ denotes $d/dr$ and $m=m(r)$ is the accumulated mass of the star.  For a given EoS, the Tolman-Oppenheimer-Volkoff~(TOV) equation can be solved as the leading order of the Einstein's field equations.  The resulting solutions are taken as the star profile to be used to  subsequently solve for the perturbation $H(r)$ at the next-to-leading order of the field equations. 

The differential equation of $H(r)$ would be integrated radially outward starting from the center with boundary conditions: $ H(r) = a_0 r^2$ and $H^{\prime}(r) = 2 a_0 r$ as $r \rightarrow 0$. The constant $a_0$ indicates how much the star is deformed due to the given external field and could be chosen arbitrarily as it cancels out in the calculation of tidal Love number. Outside the star, where $T_{\mu\nu} = 0$, the general solution for $H(r)$ at large $r$ can be written in terms of associated Legendre function $Q_{2}^2(r/M -1)\sim r^{-3}$ and $P_{2}^2(r/M -1)\sim r^{-2}$. By matching the interior and exterior solutions of $H(r)$ and their derivatives at the star surface $r=R$, this solution is uniquely determined. Equation (4) can then be used to identify the coefficients of the external solution with the axisymmetric tidal field and quadrupole moment via $\epsilon Y_{20}(\theta, \phi) = \epsilon_{i j} n^i n^j$, and $ Q Y_{20}(\theta, \phi) = Q_{i j} n^i n^j = - \lambda \epsilon_{i j} n^i n^j $, where $\epsilon $ and $Q$ are the coefficients of the $\ell = 2, m = 0$ spherical harmonic of the tidal tensor and quadrupole moment respectively. Finally, the $\ell = 2$ tidal Love number is given by~\cite{Hinderer:2009ca}

\begin{eqnarray}
k_2 &=& \frac{8 C^5}{5}\left(1-2C\right)^2\left[2+2C(y-1)-y\right]
\nonumber\\&\times&\{2C\left[6-3y+3C(5y-8)\right]
\nonumber\\&+&4 C^3\left[13-11y+C(3y-2)+2C^2(1+y)\right]
\nonumber\\&+&3(1-2C)^2\left[2-y+2C(y-1)\right]\text{ln}(1-2C)\}^{-1},
\nonumber\\\label{k2_Love}
\end{eqnarray}
where the star's compactness parameter $C \equiv M/R$ and the quantity $y \equiv R H^\prime(R)/H(R)$, which can be obtained by numerical integration of Eq.~(\ref{H_perturb}) in the region $0 < r \leq R$.

\begin{table}
\footnotesize
\begin{tabular}{ |c|c|c|c|c| }
\hline
&&&&\\
EoS & $M(M_{\odot})$ & $R$(km)& $k_2$ & $\Lambda$\\
&&&&\\
\hline
&&&&\\
& 1.8 & 14.1 & 0.106 & 302\\
&&&&\\
& 1.9 & 14.2 & 0.100 & 255\\
mq+CET, $\epsilon_s = 26$&&&&\\
& 2.0 & 13.8 & 0.0752 & 113\\
&&&&\\
& 2.2 & 12.2 & 0.0285 & 14.0\\
&&&&\\
\hline
&&&&\\
& 1.8 & 12.8 & 0.0619 & 108\\
&&&&\\
mq+CET, $\epsilon_s = 28$ & 1.9 & 12.3 & 0.0453 & 50.0\\
&&&&\\
& 2.0 & 11.8 & 0.0328 & 22.7\\
&&&&\\
\hline
&&&&\\
& 1.5 & 12.0 & 0.100 & 313\\
mq+FYSS, $Y = 0.10$&&&&\\
$\epsilon_s = 23.2037$& 1.9 & 11.9 & 0.0577 & 51.8\\
&&&&\\
\hline
&&&&\\
& 1.5 & 11.9 & 0.0956 & 290\\
mq+FYSS, $Y = 0.14$&&&&\\
$\epsilon_s = 23.2037$& 1.9 & 11.8 & 0.0550 & 48.3\\
&&&&\\
\hline
\end{tabular}
\caption{Tidal love number $k_2$ and dimensionless tidal deformation $\Lambda$ of NS with multiquark core and nuclear CET/FYSS crust.}
\label{tabI}
\end{table}

\begin{table}
\footnotesize
\begin{tabular}{ |c|c|c|c|c|c| }
\hline
&&&&&\\
EoS &$M_1(M_{\odot})$ &$M_2(M_{\odot})$ &$\Lambda_1$& $\Lambda_2$&$\tilde{\Lambda}$\\
&&&&&\\
\hline
&&&&&\\
&1.7 & 1.7 & 418 & 418 & 418 \\
&&&&&\\
mq+CET, $\epsilon_s = 26$ & 2.0 & 1.4 & 113 & 1130 & 391\\
&&&&&\\
& 1.87 & 0.93 & 247 & 7980 & 1280\\
&&&&&\\
\hline
&&&&&\\
&1.7 & 1.7 & 260 & 260 & 260 \\
&&&&&\\
mq+CET, $\epsilon_s = 28$ & 2.0 & 1.4 & 22.7 & 1120 & 313\\
&&&&&\\
& 1.87 & 0.93 & 62.3 & 8080 & 1080\\
&&&&&\\
\hline
&&&&&\\
&1.7 & 1.7 & 128 & 128 & 128 \\
&&&&&\\
mq+FYSS, $Y = 0.10$, & 2.0 & 1.4 & 32.2 & 498 & 158\\
$\epsilon_s = 23.2037$&&&&&\\
& 1.87 & 0.93 & 59.5 & 5670 & 776\\
&&&&&\\
\hline
&&&&&\\
&1.7 & 1.7 & 118 & 118 & 118 \\
&&&&&\\
mq+FYSS, $Y = 0.14$, & 2.0 & 1.4 & 30.2 & 459 & 146\\
$\epsilon_s = 23.2037$&&&&&\\
& 1.87 & 0.93 & 55.4 & 5310 & 727\\
&&&&&\\
\hline
\end{tabular}
\caption{Dimensionless tidal deformation $\Lambda_1$, $\Lambda_2$ and combined dimensionless tidal deformation $\tilde{\Lambda}$ of binary systems of NS with multiquark core and nuclear CET/FYSS crust.}
\label{tabII}
\end{table} 
For the binary system, each mass is deformed by companion star's gravitational field, the combined deformation parameter of the binary $M_{1}, M_{2}$ is defined as

\begin{equation}
\tilde{\Lambda} \equiv \frac{16}{13} \frac{(M_1+12 M_2)M_1^4 \Lambda_1 + (M_2+12 M_1)M_2^4 \Lambda_2}{(M_1 + M_2)^5}. \notag
\label{clamda_dimless}
\end{equation}

The value of $k_{2}$ for MQS and NS with multiquark core is plotted with respect to mass in Fig.~\ref{k2fig}.  The dimensionless $\Lambda$ is plotted in Fig.~\ref{lamMfig}.  We consider the MQS with $\epsilon_{s}=23.2037, 26, 28$ GeV/fm$^{3}$ and the NS with multiquark core continuing to nuclear CET~(FYSS) crust for $\epsilon_{s}=26, 28~(23.2037)$ GeV/fm$^{3}$.  In Fig.~\ref{lamMfig}, the Gravitational Waves events detected by LIGO/Virgo, GW170817 and GW190425, are also presented.  The lower and upper mass limits of both the binary masses are given by ($1.17 M_{\odot}, 1.60 M_{\odot}$) for GW170817 at 90\% confidence interval.  This interval is converted to $1-\sigma$ or 68.27\% confidence interval ($1.25 M_{\odot}, 1.52 M_{\odot}$) in Fig.~\ref{lamMfig}. Similarly for GW190425, the 90\% confidence interval ($1.46 M_{\odot}, 1.87 M_{\odot}$) becomes ($1.54 M_{\odot}, 1.79 M_{\odot}$) for $1-\sigma$ 68.27\% confidence interval.  Constraints on $\Lambda$ for each gravitational event~\cite{Abbott:2020uma,TheLIGOScientific:2017qsa} are presented as upper limits in Fig.~\ref{lamMfig}, i.e., $\Lambda < 800$ for GW170817 and $\Lambda < 600$ for GW190425.  Table~\ref{tabI} shows $k_{2}, \Lambda$ for NS with CET/FYSS crust and multiquark core for certain large masses relevant to detections of Gravitational-wave Observatories~\cite{Abbott:2020uma,TheLIGOScientific:2017qsa,GWcollab}.  Table~\ref{tabII} explores the combined $\tilde{\Lambda}$ for binaries with masses possibly in the mass range constrained by LIGO/Virgo~\cite{Abbott:2020uma,TheLIGOScientific:2017qsa}.  The loose constraints are $\tilde{\Lambda}< 600, 700-800$ for the binary with masses $(1.5-1.7, 1.6-1.9)M_{\odot}$~(GW190425, low-spin prior) and $(0.9-1.4, 1.4-2.3)M_{\odot}$~(GW170817) respectively.  We choose the mass range of each companion in the binary to be larger than $1.7M_{\odot}$ so that it has multiquark core.

\begin{figure}[h]
	\centering
	\includegraphics[width=0.48\textwidth]{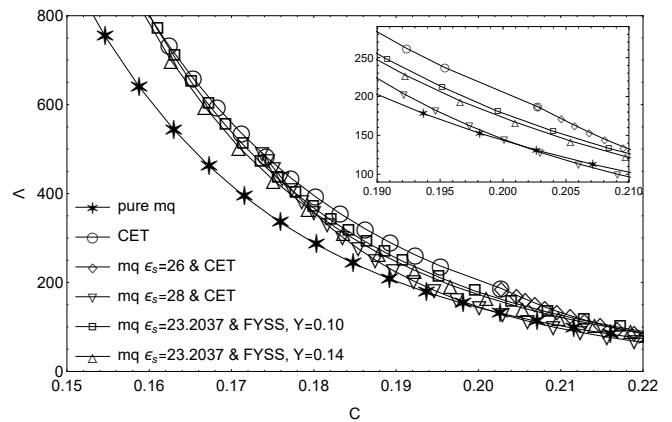}
	\caption{Dimensionless tidal deformation $\Lambda$ versus compactness $C=M/R$ of MQS/NS with MQ core.}
	\label{LamCfig}
\end{figure}

Yagi and Yunes discovered separate universal classes of compact stars with nuclear and quark matter EoS using the characteristic relationship between the moment of inertia, (rotational) Love number and the quadrupole moment~\cite{Yagi:2013bca}.  It is interesting to see which class the NS with multiquark core/MQS would fit into.  Instead of using the rotational deformability parameter, we plot relationship between $\Lambda$ and the compactness $C\equiv M/R$ in Fig.~\ref{LamCfig}.

Similar to the rotational deformation case~(between typical nuclear NS and strange quark stars), two separate deformability classes of pure MQS and pure nuclear NS curves exist as shown in Fig.~\ref{LamCfig}.  Pure MQS has lower $\Lambda$ at a given $C$ than the pure nuclear NS and the hybrid star interpolates between the two classes.  The inverted triangle and diamond plot in Fig.~\ref{LamCfig} show the NS with MQ core transiting from pure MQS curve to the CET curve as $C$ decreases.  NS with high-$Y(\equiv n_{p}/n_{b}=0.10, 0.14$, where $n_{p}~(n_{b})$ is the number density of proton~(baryon) respectively) FYSS crust sides more with the nuclear CET neutron star curve.  Remarkably due to the independence of $\epsilon_{s}$ of the compactness $C$,  all MQSs with any $\epsilon_{s}$ degenerate on the same curve.

\section{Linearized adiabatic radial pulsations of the holographic multiquark star}  \label{secpul}

Infinitesimal radial pulsations of star were first discussed under general relativity framework by Chandrasekhar~\cite{Chandrasekhar:1964zz}.  Following the rewritten form in Ref.~\cite{Chanmugam,Gondek:1997fd}, there are 2 crucial quantities describing the infinitesimal radial pulsations: the relative radial displacement, $\xi = \Delta r/r$, 
where $\Delta r$ is the radial displacement of the matter content of the star, and $\Delta P$, the corresponding Lagrangian perturbation of the pressure.  The two essential differential equations governing the radial pulsation can be expressed as

\begin{eqnarray}
\frac{d\xi}{dr} &=& -\frac{1}{r}(3\xi +\frac{\Delta P}{\Gamma P}) -\frac{dP}{dr}\frac{\xi}{(P+\rho )},\label{Delxi}\\
\frac{d\Delta P}{dr} &=& \xi\left\{\omega^2 r e^{\lambda-\nu}(P+\rho ) -4 \frac{dP}{dr}\right\} \nonumber\\
&+& \xi\left\{\left(\frac{dP}{dr}\right)^2 \frac{r}{(P+\rho)} -8\pi r e^\lambda(P+\rho )P  \right\} 
\nonumber\\
&+& \Delta P\left\{\frac{dP}{dr} \frac{1}{(P+\rho)} -4\pi r e^\lambda (P+\rho)  \right\}, \label{DelP}
\end{eqnarray}
where $\Gamma$ is the adiabatic index defined in Section~\ref{secgam}, $\omega$ is the eigenfrequency of the radial oscillation and the $\xi$ and $\Delta P$ are assumed to have a harmonic time dependence $\propto e^{-i\omega t}$. A key feature of this system of radial pulsation equations, Eqs.~(\ref{Delxi}) and (\ref{DelP}), is that they do not involve any derivatives of the adiabatic index, $\Gamma$. 

There is subtlety in the calculation of second-order perturbation (\ref{DelP}) involving $\nu(r)$ from the zeroth-order star profile.  We need to choose $\nu(r=0)$ such that 
\be
e^{\nu(R)}=1-\frac{2M}{R},  \label{nueq}
\ee
at the star surface.  Even though the Einstein's field equation has shift symmetry under $\nu\to \nu+\nu_{0}$ for arbitrary constant $\nu_{0}$ at the zeroth order, the gauge fixing of $\nu$ according to (\ref{nueq}) is required for the calculation of first and second-order perturbations.  This is the matching condition to the Schwarzschild metric in the exterior region of the star.  For $\lambda(r)$, we have used the substitution
\be
e^{-\lambda(r)}=1-\frac{2m(r)}{r},
\ee
in every perturbation equation so that the matching to the Schwarzschild metric is thus guaranteed. 

To solve the two coupled linear differential equations, Eqs.~(\ref{Delxi}) and (\ref{DelP}), the boundary conditions at the center and surface of the star must be specified. The first natural condition is the coefficient of the $1/r$-term in Eq.~(\ref{Delxi}) must vanish to avoid divergence as $r \to 0$,

\begin{eqnarray}
\Delta P(r=0) = -3 (\xi \Gamma P)|_{r=0}. \label{BDelP0}
\end{eqnarray}
The relative radial displacement is normalized to unity at the center, $\xi(0) = 1$. The surface of the star $R$ is defined by $P(r=R)=0$. This also leads to
\begin{eqnarray}
\Delta P(r=R) = 0. \label{BDelPR}
\end{eqnarray}
After solving the TOV for the star profile for a given EoS, the two coupled linear equations can be subsequently solved by the shooting method for the eigenvalue $\omega^2$ and the perturbation profile $\xi (r)$ and $\Delta P(r)$ satisfying the boundary conditions, Eqs.~(\ref{BDelP0}) and (\ref{BDelPR}). For a given EoS model of static star, a set of the eigenvalues; $ \omega_0^2 < \omega_1^2 < \cdots < \omega_{n}^2 < \cdots $, with corresponding eigenfunctions;  $\xi_0, \xi_1, \cdots, \xi_{n}, \cdots,$ could be obtained.  The characteristic standing-wave eigenfunction $\xi_{n}$ has $n$ nodes within the star.  The eigenfunctions of pressure perturbation $\Delta P_{n}(r)$ are simultaneously obtained.

For $\omega_{n}^2 > 0$ for all $n$, the static stellar model is stable with respect to small radial adiabatic perturbations.  If the lowest eigenfrequency $\omega_0 = 0$, the configuration is on the verge of stability/instability~\cite{Cox1980}.  Notably, the star configuration can be stable with respect to a set of higher-modes pulsations and become unstable for lower modes.  In this work, only the zeroth-mode instability is studied.  

\subsection{Eigenfrequencies and instabilities}

\begin{figure}[h]
	\centering
	\includegraphics[width=0.48\textwidth]{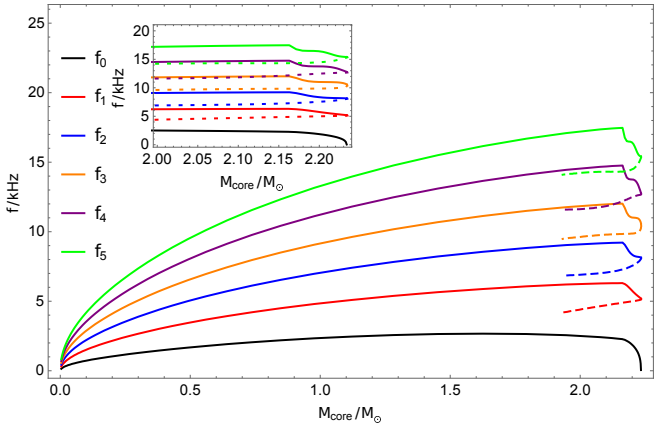}
	\caption{\raggedright{The eigenfrequencies $f_{n}=\omega_{n}/2\pi$ of $n = 0, 1, 2, 3, 4, 5$ modes versus mass of multiquark core at energy density scale $\epsilon_s = 23.2037 $ GeV/fm$^{3}$. Solid~(Dashed) line represents the frequency of the MQ core in the stable~(unstable) branch.}}
	\label{fmfig}
\end{figure}

\begin{figure}[h]
	\centering
	\includegraphics[width=0.48\textwidth]{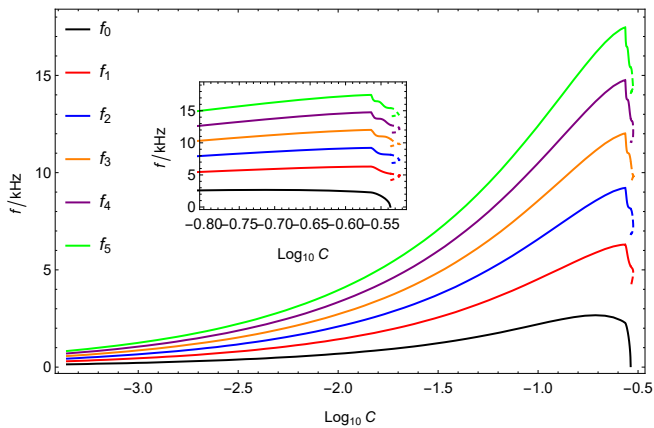}
	\caption{\raggedright{The eigenfrequencies $f_{n}=\omega_{n}/2\pi$ of $n = 0, 1, 2, 3, 4, 5$ modes versus compactness $C=M/R$ of multiquark core at energy density scale $\epsilon_s = 23.2037 $ GeV/fm$^{3}$. Solid~(Dashed) line represents the frequency of the MQ core in the stable~(unstable) branch.}}
	\label{fcfig}
\end{figure}

According to the radial pulsation Eqs.~(\ref{Delxi}) and (\ref{DelP}) together with the TOV equation, the eigenfrequencies could be determined numerically by the shooting process and they have been plotted in Fig.~\ref{fmfig} and Fig.~\ref{fcfig}.  In the SS model, the frequencies of the multiquark-core oscillations scale with $\sqrt{\epsilon_{s}}$.  The zeroth-mode frequency is about $2.5$ kHz for massive NS with masses around $2M_{\odot}$ for $\epsilon_{s}=23.2037$ GeV/fm$^{3}$.  Instability {\it with respect to the zeroth mode} appears when the central density exceeds the critical value at the maximum mass and enters into the unstable branch of the $M-R$ diagram as expected.  For NS with masses above $2.15 M_{\odot}$~(for $\epsilon_{s}=23.2037$ GeV/fm$^{3}$), higher-mode oscillations show distinct wiggly pattern of frequencies with respect to the mass due to the interplay between perturbing waves in the ``mqh'' and ``mql'' structures. 


\section{Conclusions and Discussions}\label{sec-con}

The deformation parameters of the NS with multiquark core~(or hybrid star) and MQS have been calculated in the holographic SS model.  With the nuclear CET and FYSS crust EoS, the massive NS with multiquark core is consistent with observational constraints from the LIGO/Virgo on the mass, radius~\cite{Pinkanjanarod:2020mgi}, and deformation parameter $k_{2},\Lambda$.  Deformability of MQ-matter star is clearly distinguishable from the baryonic-matter star when plotted against the compactness $C$.  Hybrid stars such as the massive NS with MQ core has the deformability $\Lambda-C$ profile that interpolates between the pure MQS and the pure baryonic NS.  

Characteristic eigenfrequencies of the radial pulsation of the multiquark core are calculated for the entire mass range for $n=0-5$ modes.  The fundamental mode frequency is approximately $2.5$ kHz near the maximum masses around $2M_{\odot}$.  A unique characteristic of the fundamental frequency is that it is slowly increasing for a wide mass range up until approaching the maximum mass where it drops abruptly to small value as shown in Fig.~\ref{fmfig}.  For $M_{\rm core}$ close to the maximum mass, the higher modes have more wiggly pattern which is originated from the interplay between radial waves in the high~(``mqh'') and low~(``mql'') density region of the MQ core.

The distinctive pattern in $\Lambda$ vs. mass~(and compactness) where there is kink at the transition between MQ core and nuclear crust could help identifying the existence of MQ core in the massive NS.  The sets of pulsation frequencies of the MQ core can be compared with observations and EoS of other models and reveal the properties of nuclear matter at extreme densities.  For example, whether the multiquarks could form and remain in the fluid/gas phase or diquark could form superfluid below certain CSC temperature or both kinds of multiquark phases, MQ and CSC, could be found under the appropriate situations inside the NS~(see e.g. Ref.~\cite{BitaghsirFadafan:2018iqr,BitaghsirFadafan:2020otb} for holographic approach to CSC where the tidal deformability is also computed).  It is also possible that MQ is prefered in the most dense and warm temperature region in the core, then in the less dense and cooler layer CSC phase is prefered, and subsequently the confined nuclear is prefered in the crust.  Constraints from the observation of massive NS around $2M_{\odot}$ suggest that in such case CSC EoS should be relatively comparable to the MQ EoS.  It would be interesting to explore such possibility of NS with mixed phase in the future work.

\begin{acknowledgments}

S.P.~(first author) is supported in part by the Second Century Fund: C2F PhD Scholarship, Chulalongkorn University.  S.P. is supported by Rachadapisek Sompote Fund for Postdoctoral Fellowship, Chulalongkorn University. 
	
\end{acknowledgments}

\section{Appendix}

\begin{table}[ht]

                \footnotesize

                \begin{tabular}{ |c|c|c| }

                                \hline

                                &&\\

                                quantity & dimensionless variable & physical variable\\

                                &&\\

                                \hline

                                &&\\

                                pressure & $P$ & $\epsilon_s P$\\

                                &&\\

                                density & $\rho$ & $\displaystyle{\frac{\epsilon_s}{c^2} \rho}$ \\

                                &&\\
                                chemical potential & $\mu$ & $\epsilon_{s}\mu$~fm$^{3}$ \\
                                
                                &&\\
                                number density & $n$ & $n$~fm$^{-3}$ \\

                                &&\\

                                mass & $M$ & $\displaystyle{\frac{c^4}{\sqrt{G^3 \epsilon_s}} M}$\\

                                &&\\

                                radius & $r$ & $\displaystyle{\frac{c^2}{\sqrt{G \epsilon_s}}r}$                \\

                                &&\\

                                angular frequency & $\omega$ & $\displaystyle{\frac{\sqrt{G \epsilon_s}}{c}\omega}$ \\

                                &&\\

                                \hline

                \end{tabular}

                \caption{Conversion table of physical quantities for $\epsilon_{s}$ in GeVfm$^{-3}$ units.}

                \label{tabI}

\end{table}

\end{document}